\documentclass[a4paper,11pt]{article}
\pdfoutput=1 

\usepackage{jheppub} 

\usepackage[T1]{fontenc} 

\newcommand{\R}{\mathbb R}

\newcommand{\dif}{\mbox{d}}
\usepackage{color}
\usepackage{url}

\begin{document} 
	
\subheader{\hfill {\rm ICCUB-20-009}}

\title{\boldmath  
 Space-time Schr\"odinger symmetries of a post-Galilean particle
 }
 
\author[1]{Carles Batlle,}
\author[2]{Joaquim Gomis}

\affiliation[1]{Departament de Matem\`atiques and IOC, 
	Universitat Polit\`ecnica de Catalunya\\
	EPSEVG, Av. V. Balaguer 1, E-08800 Vilanova i la Geltr\'u, Spain}
\affiliation[2]{Departament de F\'isica Qu\`{a}ntica i Astrof\'isica and Institut de Ci\`{e}ncies del Cosmos (ICCUB), Universitat de Barcelona, Mart\'i i Franqu\`{e}s 1, E-08028 Barcelona, Spain}

\abstract{We study the space-time symmetries of the actions obtained by expanding the action for a massive free relativistic particle around the Galilean action \cite{Gomis:2019sqv}. 
We obtain all the point space-time symmetries of the post-Galilean actions by working in canonical space.
We also construct an infinite collection of generalized Schr\"odinger algebras parameterized by an integer $M$, with $M=0$ corresponding to the standard Schr\"odinger algebra. We discuss the Schr\"odinger equations associated to these algebras, their solutions and projective phases.}

\emailAdd{carles.batlle@upc.edu}
\emailAdd{joaquim.gomis@ub.edu}


\maketitle

\flushbottom
 \section{Introduction}
{{}The nonrelativistic gravity theory of Newton is only invariant under the Galilei group. Corrections to the
non-relativistic Newtonian theory at higher order in $1/c$ are famously important for the original experimental evidence for general relativity 
\cite{Dautcourt64,Dautcourt90}.
Higher-order (parametrized) post-Newtonian
corrections to the Keplerian two-body motion are also
of central importance in current investigations of binary
gravitational wave sources \cite{Blanchet:1995ez,Buonanno:1998gg,Blanchet:2013haa,Will:2014kxa,Damour:2015isa}. The effective one-body approach \cite{Buonanno:1998gg,Maheshwari:2016edp} was
inspired in part by the developments of the classical mechanical relativistic two-body problem \cite{Todorov:1976,PhysRevD.18.1881,PhysRevD.19.702,Giachetti:1981gr}.

Starting from the relativistic Poincar\'e algebra, one can perform an algebra contraction to the non-relativistic Galilei algebra. Alternatively, starting from the Galilei transformations one can include relativistic corrections at every order in $1/c$. With corrections up to a finite order, this system has neither Galilei nor Poincar\'e symmetry. Only when including an infinite set of corrections, with specific coefficients, does one regain the Poincar\'e symmetry. However, it has recently been argued that the finite-correction-case has a symmetry algebra
\cite{Gomis:2019sqv}, but this requires enlarging the space on which the transformations act.
}

 It is well known that the most general point symmetries of the Schr\"odinger equation and the free non-relativistic particle is the Bargmann group~\cite{Bargmann:1954gh}
 with two extra generators,
the dilations and the special conformal transformations in one dimension, also called expansions~\cite{Niederer:1972zz,Hagen:1972pd}.
Recently,   the action and symmetries of a post-Galilean particle,
 which includes corrections of arbitrarily high order in $1/c$ to the non-relativistic particle  in an infinite extended Minkowski space, has been constructed \cite{Gomis:2019sqv}. This study is motivated by the importance of
higher-order post-Galilean corrections to the Keplerian two-body motion 
in current investigations of binary gravitational wave sources~\cite{Blanchet:1995ez,Buonanno:1998gg,Blanchet:2013haa,Will:2014kxa,Damour:2015isa}.

The action of a post-Galilean particle in an infinite dimensional Minkowski space 
with coordinates  $(t_{(m)}, x^a_{(m)})$, $a=1,\ldots,d$,
is obtained expanding the action for a massive free relativistic particle $S=-mc \int\dif\tau \sqrt{-\dot{X}^\mu \dot{X}_\mu}$ in powers of $1/c^2$ where the coordinates
${X}^\mu $ are\footnote{{{}The expansion is in  powers of $1/c^2$ because we are only interested in symmetries of a free theory in flat space-time and this is how $c$ appears in a post-Galilean expansion of the boosts.
In theories that also have
gravity, the expansion in $1/c$ could take into account
	strong gravity effects, see \cite{Dautcourt:1996pm,Ergen:2020yop,VandenBleeken:2017rij}.}
 }
\begin{align}
\frac{1}{c} X^0 &=   t_{(0)} + \frac{1}{c^2} t_{(1)} + \frac{1}{c^4} t_{(2)}+\ldots,\\
 X^a &=   x_{(0)}^a + \frac{1}{c^2} x_{(1)}^a + \frac{1}{c^4} x_{(2)}^a + \ldots. 
 \end{align} 
 One gets a series $S_{(0)}+ S_{(1)}+S_{(2)}+\ldots$, with $S_{(n)}$ corresponding to power $c^{2-2n}$. The first contributions are   \cite{Gomis:2019sqv}
 \begin{align}
 S_{(0)} &= -mc^2 \int\dif\tau\  \dot{t}_{(0)},\\
 S_{(1)} & = m \int\dif\tau \left(
 -\dot{t}_{(1)} + \frac{1}{2} \frac{\dot{x}_{(0)}^2}{\dot{t}_{(0)}}
 \right),\\
 S_{(2)} &= \frac{m}{c^2} \int\dif\tau \left(
 -\dot{t}_{(2)}+ \frac{\dot{x}_{(0)}^a\dot{x}_{(1)a}}{\dot{t}_{(0)}} - \frac{\dot{t}_{(1)} \dot{x}_{(0)}^2}{2\dot{t}_{(0)}^2}+ \frac{\dot{x}_{(0)}^4}{8\dot{t}_{(0)}^3}
 \right),
 \label{S2}
 \end{align}
  where $\dot{x}_{(0)}^4=(\dot{x}_{(0)}^2)^2$. 

   As shown in \cite{Gomis:2019sqv},  the symmetries of action $S_{(M+1)}$, $M\geq 0$,    realize
  the algebra~\cite{Khasanov:2011jr,Hansen:2019vqf,Gomis:2019fdh}
  \begin{align}
 \left[
 J^{n}_{ab}, J^{m}_{cd}
 \right] &= \delta_{cb}J^{(n+m)}_{ad}- \delta_{ac}J^{(n+m)}_{bd}+ \delta_{bd}J^{(n+m)}_{ac}- \delta_{bd}J^{(n+m)}_{ac},\nonumber\\
 \left[
 J^{n}_{ab}, B^{m}_{c}
 \right] &= \delta_{bc} B^{(n+m)}_a - \delta_{ac} B^{(n+m)}_b,\nonumber\\
  \left[
  J^{n}_{ab}, P^{m}_{c}
  \right] &= \delta_{bc} P^{(n+m)}_a - \delta_{ac} P^{(n+m)}_b,\nonumber\\
  \left[
  B^{(n)}_a, P^{(m)}_b
  \right] &= \delta_{ab}H^{(n+m+1)},\nonumber\\
    \left[
    H^{(m)},B^{(m)}_a
    \right] &= -P^{(n+m)}_a,\nonumber\\
    \left[
     B^{(n)}_a, B^{(m)}_b
     \right] &= J^{(n+m+1)}_{ab},
     \label{KMGal}
 \end{align}
  truncated at level $M$, that is, for $n,m=0,1,\ldots,M$, with any generator with index higher than $M$ set to zero, except for $H^{(M+1)}$   which plays  the role of a central charge of the truncated algebra.\footnote{From the abstract algebra point of view, one could also have the non-central charges $J^{(M+1)}_{ab}$, but they are not realized in the actions $S_{(M+1)}$.} 
The generators $B^{(0)}_a$ and $J^{(0)}_{ab}$ generate, respectively, Galilean boosts and ordinary rotations, while  $B^{(n)}$ and $J^{(n)}_{ab}$ for $n\geq 1$ are generalized boosts and rotations which transform coordinates between different levels of the expansion in $1/c^2$.

 Disregarding the total derivative term $-m/c^{2M} \dot{t}_{(M+1)}$ which appears in $S_{(M+1)}$, $M\geq 0$,   the canonical analysis of the  above actions suggest 
 the presence of $M+1$ first class primary constraints \cite{Gomis:2019sqv}
 and therefore the existence of $M+1$ gauge transformations.
 For instance, for $M=0$ one has the single constraint
 	\begin{align}
 	\phi^{(0)} &= \frac{1}{2} p_a^{(0)} p^{(0)a}- m E^{(0)},
 	\label{constraints0}
 	\end{align}
 	while for $M=1$ there are two constraints\footnote{Here and later on, the constraints and the generators that we will define should be understood as having an additional subscript denoting the value of $M$. We prefer not to write it in order to not to charge the notation too much. Since all the computations are done for fixed $M$, this should not cause any confusion.}
 	\begin{align}
 	\phi^{(0)} &= p_a^{(0)}p^{(1)a} - \frac{1}{2} E^{(1) 2} - \frac{m}{c^2}E^{(0)},\nonumber\\
 		\phi^{(1)} &= \frac{1}{2} p_a^{(1)} p^{(1)a}- \frac{m}{c^2} E^{(1)}.
 		\label{constraints1}
 	\end{align}
 
 Here $p^{(0)}_a$ and $p^{(1)}_a$ are the canonical momenta associated to $x_{(0)a}$ and $x_{(1)a}$, respectively,   where
 $E^{(0)}$ and $E^{(1)}$ are minus the canonical conjugated  momenta of $t_{(0)}$ and $t_{(1)}$.

  The presence of the above $M$  first class constraints means that there are $M$ gauge symmetries. Fixing the gauge does not destroy the space-time invariances of the theory and just introduces a different realization. 
  It should be noticed \cite{Gomis:2019sqv,Gomis:2020GC} that, if in addition to fixing the gauge, one performs an adequate projection in the space of all the $x$, each of the actions in this sequence yields \textit{all} the terms of the action obtained by expanding the relativistic action up to the given order. 
 
As is well known~\cite{Niederer:1972zz,Hagen:1972pd}, for $M=0$ the Galilean algebra can be extended to the Schr\"odinger algebra adding two new generators, $D$, the generator of space-time dilatations, and $C$, which generates  special conformal transformations. Together with the Hamiltonian $H$, they form a $sl(2,\R)$ subalgebra which, in the normalization for $D$ that we will use, reads
$\{D,C\} =-2C,\ \{D,H\}=2H,\ \{C,H\}=D$.

In this paper we first construct the explicit form of the $M+1$ of the mass-shell 
constraints associated to the action $S_{(M+1)}$, $M\geq 0$. These constraints 
are first class and allow 
us to construct the canonical action $S_{M+1}^c.$ 
We write  then the most general canonical generator  $G$ linear in the momenta.
The request that $G$  be a constant of motion implies a set of partial differential equations 
for the unknown functions of the  generator $G$ that we solve\footnote{
For the use of this method in non-relativistic systems see for example 
\cite{Cariglia:2014dwa,Batlle:2016iel} and references therein.}.
This allows to construct the most general point symmetry transformation of the
canonical action $S_{M+1}^c$.

 The algebra of  these transformations is a generalization of the Schr\"odinger algebra,
 and does not contain, except for the case $M=0$, an $sl(2,\R)$ algebra, 
 and therefore they are not truncations of the  Schr\"odinger-Virasoro algebra \cite{Roger:2006rz}, and they are also different from the conformal Galilean algebras (CGA) with 
  dynamical exponent $z=2/N$, with $N$ positive integer, since these contain an $sl(2,\R)$
  subalgebra \cite{henkel1997local,negro1997nonrelativistic,Duval:2009vt,Duval:2011mi}.
 The algebras that we obtain in this paper  contain the generators $H^{(n)}, B^{(n)}, J^{(n)}_{ab}$ that close 
 under the algebra (\ref{KMGal}) and the new generators $D^{(n)}$, which generalize the dilatations,
  and a single generator $C$ of expansions.

 We  consider the $M+1$ Sch\"odinger equations associated to the quantization 
 of the post-Galilean particle described by the action $S_{M+1}^c$. We also study the
 projective character of the wave function.

  The paper is organized as follows. In Section \ref{can_act} we write the canonical action $S^c_{M+1}$ for arbitrary $M$. In Section \ref{postN2} we deal with the $M=1$ case, for which we compute the most general point transformation by means of the canonical version of Noether's theorem  and obtain the first extended Schr\"odinger algebra. Section \ref{genSA} contains our main results, and we present the algebra for arbitrary $M$. 
   The Schr\"odinger equations associated to the new algebras and their projective invariance are discussed in Section \ref{GS_PI}.
    Our results  are summarized in Section \ref{conclusions}, and some open problems are discussed.  Appendix \ref{inv_constraints} contains the proof of the invariance of the constraints under generalized boosts and rotations proposed in Section \ref{can_act}, Appendix \ref{TVC} lists the transformations of all the canonical variables, and finally Appendix \ref{IVC} discusses the invariance of the constraints  under the full set of transformations.
 
 \section{The canonical action of a post-Galilean particle}
 \label{can_act}

In this section we  construct the canonical action $S_{M+1}^c$ associated to the
post-Galilean  action $S_{(M+1)}$ for a generic $M$, written in terms of the mass-shell constraints
$\phi^{(k)}$
\begin{equation}\label{canonicalaction}
 S_{M+1}^c=\int d\tau\sum_{k=0}^{M}\big(- E^{(k)} \dot{t}_{(k)}+ p^{(k)}_a \dot{x}_{(k)}^a
 -e_{(k)} \phi^{(k)}\big).
 \end{equation}
The constraints $\phi^{(k)}$ are the generalization of those for $M=0$ and $M=1$
\cite{Gomis:2019sqv,Gomis:2020GC}.
 Explicitly, 
the $M+1$ constraints corresponding to level $M$ are
 \begin{equation}
 \phi^{(k)} = -\frac{m}{c^{2M}} E^{(k)} + \frac{1}{2} \sum_{l=k}^M p_a^{(l)} p^{(M+k-l)a} 
 - \frac{1}{2} \sum_{l=k}^{M-1} E^{(l+1)} E^{(M+k-l)},
 \label{constraintsM}
 \end{equation}
 for $k=0,1,\ldots,M$. They reproduce  (\ref{constraints0}) for $M=0$ and (\ref{constraints1}) for $M=1$, and for the next few $M$ they yield
 \begin{align}
 	\phi^{(0)} &= p_a^{(0)} p^{(2)a} + \frac{1}{2} p_a^{(1)}p^{(1)a}- E^{(1)}E^{(2)} - \frac{m}{c^4} E^{(0)},\\
 	\phi^{(1)} &= p_a^{(1)}p^{(2)a} - \frac{1}{2} E^{(2) 2} - \frac{m}{c^4}E^{(1)},\\
 	\phi^{(2)} &= \frac{1}{2} p_a^{(2)} p^{(2)a}- \frac{m}{c^4} E^{(2)}.
 	\end{align}
 for $M=2 $ and
 	\begin{align}
 	\phi^{(0)} &= p_a^{(0)}p^{(3)a} + p_a^{(1)}p^{(2)a} - \frac{1}{2} E^{(2)2} - E^{(1)}E^{(3)} - \frac{m}{c^6} E^{(0)},\\
 	\phi^{(1)} &= p_a^{(1)} p^{(3)a} + \frac{1}{2} p_a^{(2)}p^{(2)a}- E^{(2)}E^{(3)} - \frac{m}{c^6} E^{(1)},\\
 	\phi^{(2)} &= p_a^{(2)}p^{(3)a} - \frac{1}{2} E^{(3) 2} - \frac{m}{c^6}E^{(2)},\\
 	\phi^{(3)} &= \frac{1}{2} p_a^{(3)} p^{(3)a}- \frac{m}{c^6} E^{(3)}.
 	\end{align}
 for $M=3$. 
 Notice that the last $M$ constraints for level $M$ coincide in form with the $M$ constraints from level $M-1$, with different variables, and that the new constraint at each level is $\phi^{(0)}$.
 We can see that the  constraints are first class if we use the Poisson brackets
\begin{align}\label{PB}
\{E^{(k)},t_{(j)}\} = \delta^k_j,\quad &
\{x_{(k)}^a,p_b^{(j)}\} = \delta^a_b \delta_k^j, \ \ k,j=0,1,\ldots,M.
\end{align}

For each $M$, eliminating the $M+1$ momenta $p^{(k)}_a$,  energies 
 $E^{(k)}$ and    multipliers $e_{(k)}$ one can recover  the corresponding action in configuration space \cite{Gomis:2019sqv}.

Prompted by the standard Galilean boost and rotation transformations on momenta and energy
\begin{equation}
\delta p_a^{(0)} = m v_{(0)a}, \quad \delta E^{(0)}= v_{(0)}^a p^{(0)}_a,
\end{equation}
 we propose the following generalized transformations for momenta and energies
 \begin{align}
 \delta p_a^{(k)} &=  \sum_{j=0}^{M-k-1} v_{(j)a}\ E^{(k+j+1)} + \frac{m}{c^{2M}} v_{(M-k)a} +  \sum_{j=0}^{M-k} \omega_{(j)ab}\ p^{(k+j)b},\nonumber\\
 \delta E^{(k)} &=  \sum_{j=0}^{M-k} v_{(j)}^a p_a^{(k+j)} .
 \label{Ep_trans}
 \end{align}
 It can be seen that they realize  the algebra  (\ref{KMGal}) truncated at level $M$, with central extension $H^{(M+1)}=-m/c^{2M}$. 
Furthermore,  as shown in Appendix \ref{inv_constraints}, for given $M$ all the constraints $\phi^{(k)}$, $k=0,1,\ldots,M$, are invariant under these transformations of momenta and energies, and this provides a further justification for the form of the constraints (\ref{constraintsM}) for general $M$.
The kinetical term in the action
$- E^{(k)} \dot{t}_{(k)}+ p^{(k)}_a \dot{x}_{(k)}^a$  is quasi-invariant under transformations
(\ref{Ep_trans}) and the corresponding transformations for ${t}_{(k)},{x}_{(k)}^a$ (see \cite{Gomis:2019sqv} and Appendix \ref{TVC}).

\section{Symmetries of post-Galilean particle for M=1}
\label{postN2}

 The action (\ref{S2}) for the second order expansion of a post-Galilean particle in $d+1$ space-time \cite{Gomis:2019sqv,Gomis:2020GC} without the total derivative $-\frac{m}{c^2}\dot{t}_{(2)}$ becomes
\begin{equation}
S_2= \frac{m}{c^2} \int\dif\tau \left(
\frac{\dot{x}_{(0)}^a\dot{x}_{(1)a}}{\dot{t}_{(0)}} - \frac{\dot{t}_{(1)} \dot{x}_{(0)}^2}{2\dot{t}_{(0)}^2}+ \frac{\dot{x}_{(0)}^4}{8\dot{t}_{(0)}^3}
\right) = \int\dif\tau  L_2.
\label{S2b}
\end{equation}

The canonical momenta are given by
\begin{align}
E^{(0)} &= -\frac{\partial L_2}{\partial \dot{t}_{(0)}} = - \frac{m}{c^2} \left(
- \frac{\dot{x}_{(0)}^a\dot{x}_{(1)a}}{\dot{t}_{(0)}^2} + \frac{\dot{t}_{(1)}\dot{x}_{(0)}^2}{\dot{t}_{(0)}^3}- \frac{3}{8} \frac{\dot{x}_{(0)}^4}{\dot{t}_{(0)}^4}
\right),\\
E^{(1)} &= - \frac{\partial L_2}{\partial \dot{t}_{(1)} } = \frac{m}{2c^2} \frac{\dot{x}_{(0)}^2}{\dot{t}_{(0)}^2},\\
p^{(0)}_a &= \frac{\partial L_2}{\partial \dot{x}_{(0)}^a} = \frac{m}{c^2} \left(
\frac{\dot{x}_{(0)a}}{\dot{t}_{(0)}} -\frac{\dot{t}_{(1)}\dot{x}_{(0)a}}{\dot{t}_{(0)}^2} + \frac{1}{2}\frac{\dot{x}_{(0)}^2 \dot{x}_{(0)a}}{\dot{t}_{(0)}^3}
\right),\\
p^{(1)}_a &= \frac{\partial L_2}{\partial \dot{x}_{(1)}^a} = \frac{m}{c^2} \frac{\dot{x}_{(0)a}}{\dot{t}_{(0)}},
\end{align}
and they obey the primary first-class constraints
\begin{align}
\phi^{(0)} &= p_a^{(0)} p^{(1)a} -\frac{1}{2} E^{(1)2} - \frac{m}{c^2}E^{(0)},\\
\phi^{(1)} &= \frac{1}{2} p^{(1)2}- \frac{m}{c^2} E^{(1)},
\end{align}
which agree with (\ref{constraintsM}) for $M=1$.
The Dirac Hamiltonian is given by
\begin{align}
H_D &= e_{(0)} \phi^{(0)} + e_{(1)} \phi^{(1)}\nonumber\\
 &= e_{(0)} \left(
p_a^{(0)} p^{(1)a} -\frac{1}{2} E^{(1)2} - \frac{m}{c^2}E^{(0)}
\right) +
e_{(1)} \left( 
 \frac{1}{2} p^{(1)2}- \frac{m}{c^2} E^{(1)}
  \right),
\label{HD}
\end{align}
and yields the equations of motion
\begin{align}
& \dot{t}_{(0)} = -\frac{\partial H_D}{\partial E^{(0)}} = \frac{m}{c^2} e_{(0)},\quad
\dot{t}_{(1)} = -\frac{\partial H_D}{\partial E_1} = e_{(0)} E^{(1)} +\frac{m}{c^2} e_{(1)},\\
& \dot{x}_{(0)}^a = \frac{\partial H_D}{\partial p^{(0)}_a} =  e_{(0)} p^{(1)a},\quad
\dot{x}_{(1)}^a = \frac{\partial H_D}{\partial p^{(1)}_a} =  e_{(0)} p^{(0)a} + e_{(1)} p^{(1)a},\\
&\dot{E}^{(0)} = \frac{\partial H_D}{\partial t_{(0)}} =0,\quad
\dot{E}^{(1)} = \frac{\partial H_D}{\partial t_{(1)}}  =0,\\
&\dot{p}^{(0)}_a = -\frac{\partial H_D}{\partial x_{(0)}^a} =0,\quad
\dot{p}^{(1)}_a = -\frac{\partial H_D}{\partial x_{(1)}^a} =0.
\end{align}
 
\subsection*{Space-time symmetries}
The canonical generator of space-time symmetries is 
given by
\begin{equation}
G=-E^{(0)} \eta_{(0)} - E^{(1)} \eta_{(1)} + p^{(0)}_a \xi_{(0)}^a +p^{(1)}_a \xi_{(1)}^a - \delta F,
\label{Gd}
\end{equation}
with the $\eta$, $\xi$  and $\delta F$ are  unknown functions of $t_{(0)}$, $t_{(1)}$, $x_{(0)}$ and $x_{(1)}$, so that the space-time symmetries are obtained as $\delta t_{(0)} = \eta_{(0)}$, $\delta t_{(1)}=\eta_{(1)}$, $\delta x_{(0)}^a = \xi_{(0)}^a$, $\delta x_{(1)}^a = \xi_{(1)}^a$, and then $\delta L_2 = \frac \dif{\dif\tau}\delta F$.

The equation $\dot G=0$ allows one to write the following  Killing equations \cite{Cariglia:2014dwa,Batlle:2016iel}
for the space-time symmetries of $S_2$,

 
\begin{align}
& \partial_a^{0} \eta_{(0)} =0,\ \ \partial^{1} \eta_{(0)} = 0, \ \ \partial_a^{1}\eta_{(0)}=0, \label{eq1}
\\
& \partial_a^{1}\eta_{(1)} = 0,\label{eq2}
\\
& \partial^{1} \xi_{(0)}^a = 0, \ \ \partial_a^{1} \xi_{(0)}^b =0, \label{eq3}
\\
& \partial^{0} \delta F = 0, \ \ \partial^{1}\delta  F =0, \label{eq4}
\\
& \frac{m}{c^2} \partial^{1} \xi_{(1)a} = \partial_a^{1}\delta F, \label{eq5}
\\
& \partial_a^{1} \xi_{(1)b} + \partial_b^{1} \xi_{(1)a}  = \delta_{ab} \partial^{1}\eta_{(1)}, \label{eq6}
\\
& \frac{m}{c^2} \partial^{0} \xi_{(0)a} = \partial_a^{1}\delta F, \label{eq7}
\\
& \frac{m}{c^2} \partial^{0} \xi_{(1)a} = \partial_a^{0} \delta F, \label{eq8}
\\
& \partial_a^{0} \xi_{(0)b} + \partial_a^{1} \xi_{(1)b}  = \delta_{ab} \partial^{0}\eta_{(0)}, \label{eq9}
\\ 
&  \partial_a^{0} \xi_{(1)b} + \partial_a^{0} \xi_{(1)b}  = \delta_{ab} \partial^{0}\eta_{(1)}, \label{eq10}
\\ 
& \partial^{0}\eta_{(0)} = 2 \partial^{1} \eta_{(1)}, \label{eq11}
\\ 
& \partial_a^{0} \eta_{(1)} = \partial^{1} \xi_{(1)a}, \label{eq12}
\end{align} 
for $a, b=1,\ldots, d$, and with
$
\partial^{0} = \frac{\partial}{\partial t_{(0)}}, \  \
\partial^{1} = \frac{\partial}{\partial t_{(1)}}, \ \
\partial_a^{0} = \frac{\partial}{\partial x_{(0)}^a}, \ \  
\partial_a^{1} = \frac{\partial}{\partial x_{(1)}^a}.
$

These PDE can be integrated starting from the trivial ones (\ref{eq1})---(\ref{eq4}), and one gets the unique  solution
 given by
\begin{align}
\eta_{(0)}(t_{(0)}) &= \frac{4}{3}\lambda t_{(0)} + \delta_{(0)},\label{eta0}\\
\eta_{(1)}(t_{(0)},t_{(1)},x_{(0)}) &= \frac{2}{3} \lambda t_{(1)} + v_{(0)}^a x_{(0)a} + \mu t_{(0)}^2 + \lambda_{(1)} t_{(0)} + \delta_{(1)},\label{eta1}\\
\xi_{(0)a}(t_{(0)},x_{(0)}) &= \lambda x_{(0)a} + v_{(0)a} t_{(0)} + \omega_{(0)ab} x_{(0)}^b + \varepsilon_{(0)a}, \label{xii}\\
\xi_{(1)a}(t_{(0)},t_{(1)},x_{(0)},x_{(1)}) &= \frac{1}{3} \lambda x_{(1)a} + v_{(0)a} t_{(1)} + \mu t_{(0)} x_{(0)a} + v_{(1)a} t_{(0)}\nonumber\\
& + \frac{1}{2} \lambda_{(1)} x_{(0)} + \omega_{(0)ab} x_{(1)}^b + \omega_{(1)ab} x_{(0)}^b + \varepsilon_{(1)a}, \label{zetai}\\
\delta F(x_{(0)},x_{(1)}) &= \frac{m}{2c^2} \left( \mu x_{(0)}^ax_{(0)a} + 2\ v_{(0)a} x_{(1)}^a + 2\ v_{(1)a} x_{(0)}^a \right), \label{F}
\end{align}
with $\lambda$, $\lambda_{(1)}$, $\mu$, $\delta_{(0)}$, $\delta_{(1)}$, $\varepsilon_{(0)}^a$, $\varepsilon_{(1)}^a$, 
$v_{(0)}^a$, $v_{(1)}^a$, $\omega_{(0)}^{ab}$, and $\omega_{(1)}^{ab}$ arbitrary constants, where 
$\omega_{(0)ab}=-\omega_{(0)ba}$, $\omega_{(1)ab}=-\omega_{(1)ba}$.

The non-zero (or non-constant) value of $\delta F$ is associated to the fact that we dropped the total derivative $-\frac{m}{c^2} \dot{t}_{(2)}$ from the action $S_{(2)}$. Had we kept  that term, we would have obtained $\delta t_{(2)} = -\frac{c^2}{m} \delta F$ (plus an arbitrary constant corresponding to shifts in $t_{(2)}$).

The generator of the point transformations is 
\begin{align}
G =& - E^{(0)} \left(
 \frac{4}{3}\lambda t_{(0)} + \delta_{(0)}
\right)
  -E^{(1)} \left(
\frac{2}{3} \lambda t_{(1)} + v_{(0)}^a x_{(0)a} + \mu t_{(0)}^2 + \lambda_{(1)} t_{(0)} + \delta_{(1)}
\right)
\nonumber\\
& + p^{(0)a} \left(
 \lambda x_{(0)a} + v_{(0)a} t_{(0)} + \omega_{(0)ab} x_{(0)}^b + \varepsilon_{(0)a}
\right)
\nonumber\\
& + p^{(1)a} \left(
\frac{1}{3} \lambda x_{(1)a} + v_{(0)a} t_{(1)} + \mu t_{(0)} x_{(0)a} + v_{(1)a} t_{(0)}
\right.\nonumber\\
&\hspace{1.2cm}
\left. + \frac{1}{2} \lambda_{(1)} x_{(0)} + \omega_{(0)ab} x_{(1)}^b + \omega_{(1)ab} x_{(0)}^b + \varepsilon_{(1)a}
\right)
\nonumber\\
& - \frac{m}{2c^2} \left(
  \mu x_{0}^ax_{(0)a} + 2\ v_{(0)a} x_{(1)}^a + 2\ v_{(1)a} x_{(0)}^a 
 \right),
 \label{gen2d}
 \end{align}
from which the individual generators can be defined,
\begin{align}
\lambda \to\  & D = - \frac{4}{3}E^{(0)} t_{(0)} -\frac{2}{3} E^{(1)} t_{(1)} +  p_a^{(0)} x_{(0)}^a + \frac{1}{3} p^{(1)}_a x_{(1)}^a,\label{D}\\
\delta_{(0)} \to\ & H^{(0)} = -E^{(0)}, \label{H0}\\
\delta_{(1)} \to\ & H^{(1)} = -E^{(1)}, \label{H1}\\
\varepsilon_{(0)a} \to\ & P^{(0)_a} = p^{(0)}_a,\label{Pi}\\
\varepsilon_{(1)a} \to\ & P^{(1)_a} = p^{(1)}_a, \label{Qi}\\
\mu \to\ & C = -E^{(1)} t_{(0)} ^2 + t_{(0)} p^{(1)}_a x_{(0)}^a - \frac{1}{2} \frac{m}{c^2}  x_{(0)}^ax_{(0)a}, \label{C0}\\
\lambda_{(1)} \to\ & D^{(1)} = - E^{(1)} t_{(0)} + \frac{1}{2} p^{(1)}_a x_{(0)}a, \label{C1}\\ 
v_{(0)}^a \to\ &B^{(0)}_a = - E^{(1)} x_{(0)a} + t_{(0)} p^{(0)}_a + t_{(1)} p^{(1)}_a - \frac{m}{c^2} x_{(1)a},\label{Bi}\\
v_{(1)}^a  \to\ &B^{(1)}_a= t_{(0)}p^{(1)}_a - \frac{m}{c^2} x_{(0)a}, \label{Ki}\\
\omega_{(0)}^{ab} \to\ & J^{(0)}_{ab} = p^{(0)}_a x_{(0)b} - p^{(0)}_b x_{(0)a}+ p^{(1)}_a x_{(1)b} - p^{(1)}_b x_{(1)a}, \label{Jij}\\
\omega_{(1)}^{ab} \to\ &  J^{(1)}_{ab} = p^{(1)}_a x_{(0)b} -  p^{(1)}_b x_{(0)a}. \label{Sij}
\end{align}
$H^{(0)}$, $H^{(1)}$, $P^{(0)}$ and  $P^{(1)}$ generate the translations in $t_{(0)}$, $t_{(1)}$, $x_{(0)}$ and $x_ {(1)}$, respectively.
 $D$ is the generator of dilatations, $C$   is a generator of  mixing-level special conformal transformations, $D^{(1)}$ generates mixing level dilatations and $B^{(0)}$ and $B^{(1)}$ are generators of mixing-level boosts. Finally, $J^{(0)}$ generates standard rotations at all levels, while  $J^{(1)}$ rotates the $x_{(1)}$ but puts the result in the space of $x_{(0)}$. 

\subsection*{Extended space-time algebra} 
Using the Poisson brackets (\ref{PB}) for $M=1$
one can compute the dilatation weights $\Delta_X$ of the generators, $\{X,D\} = \Delta_X X$ given in Table \ref{tableDM1}.   The standard dynamical exponent, given by the quotient of the weights of $H^{(0)}$ and $P^{(0)}$, is $z=4/3$, in contrast to the $z=2$ value of the  $M=0$ case.
\begin{table}[tb]
\centering
\begin{tabular}{|c||c|c|c|c|c|c|c|c|c|c|}
\hline
 $X$ & $C$ & $H^{(0)}$ & $H^{(1)}$ & $P^{(0)}$ &  $P^{(1)}$  & $B^{(0)}$ &  $B^{(1)}$  & $D^{(1)}$ &  $J^{(0)}$  & $J^{(1)}$ \\
 \hline
 $\Delta_X$ & $2$ & $-\frac{4}{3}$ & $-\frac{2}{3}$ &  $-1$  & $-\frac{1}{3}$ & $\frac{1}{3}$ & $1$ & $\frac{2}{3}$ & $0$ & $\frac{2}{3}$ \\
\hline
\end{tabular}
\caption{Dilatation weights of the symmetry generators for $L_2$. Weights of $t_{(0)}$, $t_{(1)}$, $x_{(0)}$ and $x_{(1)}$ are the opposites of $H^{(0)}$, $H^{(1)}$, $P^{(0)}$ and $P^{(1)}$, respectively.}
\label{tableDM1}
\end{table}

The remaining non-zero brackets among generators are
\begin{align}
& 
\{ H^{(0)}, C \} = - 2 D^{(1)},\quad
\{ H^{(0)}, D^{(1)} \} = -H^{(1)},\nonumber\\
&
\{ H^{(0)}, B^{(1)}_a \} = -P^{(0)}_a,\quad
\{ H^{(0)}, B^{(1)}_a \} = -P^{(1)}_a,\\
&
\{ H^{(1)}, B^{(0)}_a \} = -P^{(1)}_a,\\
&
\{ P^{(0)}_a, C  \} = -B^{(1)}_a,\quad 
\{ P^{(0)}_a, D^{(1)}  \} = -\frac{1}{2}P^{(1)}_a,\quad
\{ P^{(0)}_a, B^{(0)}_b  \} = -\delta_{ab} H^{(1)},\nonumber\\
&
\{ P^{(0)}_a, B^{(1)}_b  \} = \frac{m}{c^2}\delta_{ab},\quad 
\{ P^{(0)}_a, J^{(1)}_{bc}  \} = \delta_{ab}P^{(1)}_c-\delta_{ac}P^{(1)}_b,\\   
&
\{ P^{(1)}_a,B^{(0)}_b \} = \frac{m}{c^2}\delta_{ab},\quad 
\{ D^{(1)}, B^{(0)}_a  \} = -\frac{1}{2} B^{(1)}_a,\quad
\{ B^{(0)}_a,B^{(0)}_b \} =  J^{(1)}_{ab},\\
&
\{ B^{(0)}_a, J^{(1)}_{bc} \} = \delta_{ab} B^{(1)}_c - \delta_{ac} B^{(1)}_b,
\end{align}
 plus the rotation algebra of $J^{(0)}$ with itself and with all the generators with vector indexes. The central extension $H^{(2)}=m/c^2$ appears in both mixing-level translation-boost brackets  $\{P^{(0)},B^{(1)}\}$ and $\{P^{(1)},B^{(0)}\}$, instead of in $\{P^{(0)},B^{(0)}\}$ as in the Schr\"odinger algebra.
 
 Notice that $D^{(1)}$ transforms $B$ into $B$, $P$ into $P$ and $H$ into $H$, but changes the level from $^{(0)}$ to $^{(1)}$ in each case. In this sense it acts like a higher order  dilatation, and we will refer to it as a generalized dilatation.
  
The generators $H^{(0)}$, $H^{(1)}$, $D$, $C$ and $D^{(1)}$ form a solvable, indecomposable 5-dimensional subalgebra. This is in contrast with the case of the Galilean particle, where $H=H^{(0)}$, $C$ and $D$ are a realization of  the semisimple algebra $sl(2,\R)$. 
 
Non-relativistic systems with higher order derivatives and with an extended phase space
$x_{(n)}$, $p^{(n)}$ with a  single dilatation and Hamiltonian
have been proposed,  see for example
\cite{Lukierski:2002ew,Stichel:2009sz,Stichel:2013kj,Gomis:2011dw}.

 \section{Generalized Schr\"odinger algebras}
 \label{genSA}
 
We could proceed now to the $M=2$ action and follow the same procedure as before. However, for higher $M$ the computations become very involved quite rapidly and, in particular, solving the resulting system of PDE associated to the conservation of $G$ requires the use of computer algebra packages. 

Instead, in order to obtain results for arbitrary $M$, we rely on the knowledge of the $M+1$ constraints presented in Section \ref{can_act} and propose a generalization of the extended generators found for $M=0$ and $M=1$. Invariance of the constraints under the extended generators, 
together with the quasi-invariance of the kinetic terms in the canonical action, justify this approach. 
The form of the generators is further validated by the closure of the Poisson bracket algebra of generators.

The  generators of symmetries of $L_{M+1}$, $M=0,1,\ldots$, that we propose are
\begin{align}
D&= \frac{1}{2M+1}\sum_{l=0}^M (2M+1-2l) p_a^{(l)}x_{(l)}^a - \frac{1}{2M+1}\sum_{l=0}^M (2M+2-2l)E^{(l)}t_{(l)},\\
D^{(k)} &= \frac{1}{2(M+1-k)} \sum_{l=k}^M (2(M-l)+1) p_a^{(l)} x_{(l-k)}^a
\nonumber\\
&\quad\quad -\frac{1}{M+1-k} \sum_{l=k}^M (M+1-l) E^{(l)}t_{(l-k)},\\
C &= - E^{(M)} t_{(0)}^2 + t_{(0)} p_a^{(M)} x_{(0)}^a - \frac{m}{2c^{2M}} x_{(0)}^a x_{(0)a},\\
B_a^{(k)} &= \sum_{l=k}^M p_a^{(l)} t_{(l-k)} - \sum_{l=k}^{M-1} E^{(l+1)}x_{(l-k)a} - \frac{m}{c^{2M}} x_{(M-k)a},\\
H^{(k)} &= -E^{(k)},\\
P_a^{(k)} &= p_a^{(k)},\\
J_{ab}^{(k)} &= \sum_{l=k}^M \left(
p_a^{(l)}x_{(l-k)b} - x_ {(l-k)a}p_b^{(l)}
\right),
\label{fullgen}
\end{align}
where $k=0,\ldots,M$ for all the families of generators, except for the $D^{(k)}$, for which $k=1,\ldots,M$. {{}One has thus $2M+3$ scalar generators $D$, $D^{(k)}$, $C$ and $H^{(k)}$, plus $2(M+1)d$ generators $P^{(k)}_a$, $B^{(k)}_a$ and $(M+1)\frac{d(d-1)}{2}$ 
generators $J^{(k)}_{ab}$, yielding a grand total of 
\begin{equation}
2M+3+(M+1)\frac{d(d+3)}{2}
\end{equation}
 generators.}

Using the Poisson brackets (\ref{PB}) one can compute the action of these generators on  the canonical variables $x_{k}$, $p^{(k)}$, $t_{(k)}$, $E^{(k)}$, $k=0,1,\ldots,M$, and the results are given in Appendix \ref{TVC}. For boosts and rotations the transformations are, after multiplying by the corresponding parameters, those given in (\ref{Ep_trans}).

That the above generators correspond to symmetries of $S_{M+1}^c$ is proved as follows. 
First one can check (see Appendix \ref{IVC}) that the constraints $\phi^{(k)}$, $k=0,1,\ldots,M$ are invariant under the above transformations, that is,
\begin{equation}
\delta^X \phi^{(k)} = \{\phi^{(k)},X\} =0
\end{equation}
for $X$ in the set $\{
D, D^{(j)}, C, H^{(j)},P_a^{(j)}, B_a^{(j)},J_{ab}^{(j)}
\}
$.

 Furthermore, using the results in Appendix \ref{TVC} and the commutation of the transformations with the derivation with respect to $\tau$, one can prove that 
$$
K = -\sum_{k=0}^M\left( E^{(k)}\dot{t}_{(k)}+ p^{(k)}_a \dot{x}_{(k)}^a\right)
$$
is also invariant, except for the transformations corresponding to the boosts and the special conformal transformation, for which\footnote{We use $\delta_a^{B,j}$ to denote the transformation under $B^{(j)}_a$. This kind of notation will also be employed for the other generators later on.}
\begin{align}
	\delta_a^{B,j} K &= \frac{\dif}{\dif\tau}\left( \frac{m}{c^{2M}} x_{(M-j)a}   \right),\\
	\delta^C K &= \frac{\dif}{\dif\tau}\left( \frac{m}{2c^{2M}}x_{(0)}^a x_{(0)a} \right). 
\end{align}
This means that the canonical Lagrangian in $S_{M+1}^c$ is not invariant under the full set of transformations, but it is quasi-invariant,
\begin{equation}
\delta^G L_{M+1} = \frac{\dif}{\dif\tau} \delta F_{(M)},
\end{equation}
with
\begin{equation}
\delta F_{(M)} = \frac{m}{2c^{2M}} \left(
\mu x_{(0)}^a x_{(0)a} + 2 \sum_{k=0}^M v_{(k)}^a x_{(M-k)a},
\right)
\end{equation}
where $\mu$ is the parameter of the special conformal transformation and the $v_{(k)}^a$ are the boosts parameters. This agrees with the known result for $M=0$ and the result for  $M=1$ obtained in this paper and, as in those cases, $\delta F_{(M)}$ is associated with the dropping of the total derivative  
$-m/c^{2M}\dot{t}_{(M+1)}$ in $S_{(M+1)}$.

This concludes the proof that our generators yield symmetry transformations of the canonical action. Furthermore, they form a closed algebra.
Indeed, the brackets of the $D$ and $D^{(k)}$ with all the generators are
\begin{align}
\{D, D^{(k)}\} &= -\frac{2k}{2M+1} D^{(k)},\ k=1,\ldots,M,\\
\{D,C\} &= -2 C,\\
\{D,B_a^{(k)}\} &= - \frac{2k+1}{2M+1} B_a^{(k)}, \ k=0,\ldots,M,\\
\{D,H^{(k)}\} &= \frac{2M+2-2k}{2M+1} H^{(k)}, \ k=0,\ldots,M,\label{DHk}\\
\{D,P_a^{(k)}\} &= \frac{2M+1-2k}{2M+1} P_a^{(k)},\ k=0,\ldots,M,\label{DPk}\\
\{D,J_{ab}^{(k)}\} &= -\frac{2k}{2M+1} J_{ab}^{(k)},\ k=0,\ldots,M,\\
\{D^{(k)},D^{(j)}\} &= (k-j) \frac{M+1-k-j}{(M+1-k)(M+1-j)} D^{(k+j)}, \ k,j=1,\ldots,M,\nonumber\\
 &\text{provided that $k+j\leq M$, zero otherwise},\\
\{D^{(k)},C\} &= 0,\ k=1,\ldots,M,
\end{align}
\begin{align}
\{D^{(k)},B_a^{(j)}\} &= -\frac{2j+1}{2(M+1-k)} B_a^{(k+j)},\ k=1,\ldots,M,\ j=0,\ldots,M,
\nonumber\\ &\text{provided that $k+j\leq M$, zero otherwise},\\
\{D^{(k)},H^{(j)}\} &= \frac{M+1-k-j}{M+1-k} H^{(k+j)},\ k=1,\ldots,M,\ j=0,\ldots,M,
\nonumber\\ &\text{provided that $k+j\leq M$, zero otherwise},\\
\{D^{(k)},P_a^{(j)}\} &= \frac{2M+1-2(k+j)}{2(M+1-k)} P_a^{(k+j)},\ k=1,\ldots,M,\ j=0,\ldots,M,
\nonumber\\ &\text{provided that $k+j\leq M$, zero otherwise},\\
\{D^{(k)},J_{ab}^{(j)}\} &= -\frac{j}{M+1-k} J_{ab}^{(k+j)},\ k=1,\ldots,M,\ j=0,\ldots,M,
\nonumber\\ &\text{provided that $k+j\leq M$, zero otherwise}.
\end{align}

  From (\ref{DHk}) and (\ref{DPk}) for $k=0$ one can see that the standard dynamical exponent {{} associated to the Galilean variables $t_{(0)}, x_{(0)}$} depends on $M$ and is given by
\begin{equation}
z_M = \frac{2M+2}{2M+1},\ \ M=0,1,2,\ldots
\label{ZM}
\end{equation}
{{}One has that $\lim_{M\to\infty} z_M=1$, but (\ref{ZM}) takes into account only a small part of the relations, and it is not obvious what this implies at the level of the algebra itself.}

As we already noticed for $M=1$,  the above relations show that $D^{(k)}$ are a higher order version of the dilatation $D$, increasing by $k$ the levels of the generators. Furthermore, if we redefine
\begin{align}
L_k &= (M+1-k) D^{(k)}, \ k=1,\ldots,M,\\
L_0 &= \frac{2M+1}{2} D,
\end{align}
one has the non-negative part 
of a truncated   Witt algebra,
\begin{equation}
\{L_n, L_m\} = (n-m) L_{n+m},\quad n,m=0,\ldots,M,
\label{witt}
\end{equation}
provided that $n+m\leq M$, and zero otherwise.	The remaining brackets are
\begin{align}
\{B_a^{(k)},B_b^{(j)}\} &= J_{ab}^{(k+j+1)},\ k,j=0,\ldots,M,
\nonumber\\ &\text{provided that $k+j+1\leq M$, zero otherwise},\\
\{J_{ab}^{(k)},J_{cd}^{(j)}\} &= \delta_{ad}J_{bc}^{(k+j)} +
\delta_{bc}J_{ad}^{(k+j)} - \delta_{ac}J_{bd}^{(k+j)} - \delta_{bd}J_{ac}^{(k+j)},\
k,j=0,\ldots,M,
\nonumber\\ &\text{provided that $k+j\leq M$, zero otherwise},\\
\{B_a^{(k)},J_{bc}^{(j)}\} &= \delta_{ab} B_c^{(k+j)} - \delta_{ac} B_b^{(k+j)},
k,j=0,\ldots,M,
\nonumber\\ &\text{provided that $k+j\leq M$, zero otherwise},\\
\{P_a^{(k)},J_{bc}^{(j)}\} &= \delta_{ab} P_c^{(k+j)} - \delta_{ac} P_b^{(k+j)},
k,j=0,\ldots,M,
\nonumber\\ &\text{provided that $k+j\leq M$, zero otherwise},
\end{align}
\begin{align}
\{H^{(k)},B_a^{(j)}\} &= - P_a^{(k+j)},
k,j=0,\ldots,M,
\nonumber\\ &\text{provided that $k+j\leq M$, zero otherwise},\\
\{P_a^{(k)},B_b^{(j)}\} &= -\delta_{ab} H^{(k+j+1)} +  \frac{m}{c^{2M}}\delta_{ab} \delta^{k+j}_M,k,j=0,\ldots,M,
\nonumber\\ &\text{provided that $k+j+1\leq M$, zero otherwise},\\
\{C,B_a^{(k)}\} &= 0, \ k=0,\ldots,M,\\
\{C,J_{ab}^{(k)}\} &=0, \ k=0,\ldots,M,\\
\{C,P_a^{(k)}\} &=  \delta^k_0 B_a^{(M)}, \ k=0,\ldots,M,\\
\{C, H^{(k)}\} &= \begin{cases}
D & \text{if $M=0$ (so that $k=0$ is the only possible value)},\\
2 \delta^k_0 D^{(M)} & \text{if $M\geq 1$}, \ k=0,\ldots,M.
\end{cases} \label{CH}
\end{align}
The special behaviour of the last relation for $M=0$ is what  makes the $sl(2,\R)$ algebra of $D,C,H^{(0)}$ to appear for $M=0$, while for $M>0$ the $2M+3$ generators $C,D,D^{(k)},H^{(k)}$ form a solvable algebra (this is because  $D$ does not appear in the right-hand sides for $M>0$, $C$ disappears after the first derivation, and the $D^{(k)}$  and $H^{(k)}$ are dropped in successive derivations of the algebra). The central extension $H^{(M+1)}=m/c^{2M}$ appears in brackets between translations and boosts whose indexes add to $M$.

Although some of the generators can be redefined so that some of the structure constants become independent of the level $M$, as was done for the $D$, $D^{(k)}$, there are some structure constants for which the dependence on $M$ cannot be erased, that is, increasing $M$ not only  brings in new generators but it also changes the brackets of some of the old ones. 
{{}
The structure constants of the brackets among $H^{(k)}$, $P^{(k)}$, $B^{(k)}$ and $J^{(k)}$, that is, those of the original algebra (\ref{KMGal}), as well as those of $C$ with 
$P^{(k)}$, $B^{(k)}$ and $J^{(k)}$, do not depend on $M$ (the $1/c^{2m}$ can be absorbed into $m$). Using the $L_n$ instead of $D$, $D^{(k)}$, the remaining brackets, besides (\ref{witt}), are
\begin{align}
\{L_n, B_a^{(k)}   \} &= - \frac{2k+1}{2}B_a^{(k+n)},\label{LB}\\
\{L_n, P_a^{(k)}   \} &=  \frac{2M+1-2(n+k)}{2}P_a^{(k+n)},\label{LP}\\
\{L_n, J_{ab}^{(k)}   \} &= - k J_{ab}^{(k+n)},\label{LJ}\\
\{L_n, H^{(k)}   \} &= (M+1-(n+k))H^{(k+n)},\label{LH}\\
\{L_n, C \} &= - \delta_{n0}(2M+1)C,\label{LC}\\
\{C, H^{(k)}   \} &= 2\delta_0^k L_M.\label{CHb}
\end{align}
From this is clear that, for instance, one cannot redefine $P^{(k)}$ so that the dependence on $M$ dissapears from (\ref{LP}), and the same happens for (\ref{LH}) and (\ref{LC}), while (\ref{CHb}) has the problem that the generator in the right-hand side depends itself on $M$.
}


%
%


\section{Generalized Schr\"odinger equation and projective invariance}
\label{GS_PI}
The quantization of the systems that we have considered can be performed by imposing the canonical constraints on the physical states of the corresponding Hilbert space. For $M=1$ we have two constraints and we obtain a set of two generalized   Schr\"odinger equations,
\begin{align} 
-\frac{\partial^2 \Psi}{\partial x_{(0)a}\partial x_{(1)}^a} + \frac{1}{2} \frac{\partial^2 \Psi}{\partial t_{(1)}^2} 
-i \frac{m}{c^2}\frac{\partial\Psi}{\partial t_{(0)}} & =0,\label{Sch1}\\
-\frac{1}{2} \frac{\partial^2 \Psi}{\partial x_{(1)a}\partial x_{(1)}^a}-i \frac{m}{c^2}\frac{\partial\Psi}{\partial t_{(1)}} & =0,\label{Sch2}
\end{align}
where $\Psi(t_{(0)},t_{(1)},x_{(0)},x_{(1)})$ is the wave function of the physical state in coordinate representation. Working in $d=1$ for simplicity and looking for solutions of the form 
$$
\Psi(t_{(0)},t_{(1)},x_{(0)},x_{(1)})=\Psi_0(t_{(0)},x_{(0)})\Psi_1(t_{(1)},x_{(1)}),
$$ 
equation (\ref{Sch2}) can be solved by separation of variables with separation constant $\varepsilon$ to obtain
\begin{align}
\Psi_1(t_{(1)},x_{(1)}) &= A_+ e^{-i\left(\varepsilon t_{(1)} - \frac{\sqrt{2m\varepsilon}}{c} x_{(1)}  \right)}
+A_- e^{-i\left(\varepsilon t_{(1)} +\frac{\sqrt{2m\varepsilon}}{c} x_{(1)}  \right)}\nonumber\\
&\equiv  \Psi_+(t_{(1)},x_{(1)})+ \Psi_-(t_{(1)},x_{(1)}),
\label{psi1}
\end{align}
where $A_{\pm}$ are arbitrary constants. 

The separation constant $\varepsilon$ characterizes the dependence of the wave function on $t_{(1)}$, $x_{(1)}$, and is the common eigenvalue of the operators corresponding to $E^{(1)}$ and $\frac{c^2}{2m}P^{(1)2}$, that is,
 $i\partial_{t(1)}$ and $-\frac{c^2}{2m}\partial^2_{x(1)}$, respectively.

Each of the $\Psi_\pm(t_{(1)},x_{(1)})$ can be substituted into (\ref{Sch1}) and one obtains a first-order PDE for $\Psi_0$,
\begin{equation}
-i \left( \pm \frac{\sqrt{2m\varepsilon}}{c}  \right)
  \frac{\partial \Psi_0}{\partial x_{(0)}} - \frac{1}{2}\varepsilon^2  \Psi_0 - i \frac{m}{c^2}  \frac{\partial \Psi_0}{\partial t_{(0)}}=0,
\label{Sch3} 
\end{equation}
Equation (\ref{Sch3}) can be solved by the method of characteristics. Imposing the initial condition   $\Psi_0(0,x_{(0)}) =F(x_{(0)})$ at $t_{(0)}=0$, with $F$ an arbitrary smooth function, one obtains, 
\begin{equation}
\Psi_0(t_{(0)},x_{(0)}) = F_\pm\left(
x_{(0)} - \frac{c}{m}  \left( \pm \sqrt{2m\varepsilon}  \right)  t_{(0)}
\right)
e^{i \frac{c^2}{2m}\varepsilon^2 t_{(0)}},
\end{equation}
where $F_+$ and $F_-$ are the arbitrary functions corresponding  to the $\pm$ signs in (\ref{Sch3}). Finally, the total wave function solution to the system of Schr\"odinger equations is
\begin{align}
\Psi_\varepsilon(t_{(0)},t_{(1)},x_{(0)},x_{(1)}) &= F_+\left( x_{(0)} - \frac{c}{m}\sqrt{2m\varepsilon}\ t_{(0)}   \right)
  e^{-i\left( \varepsilon t_{(1)} - \frac{\sqrt{2m\varepsilon}}{c} x_{(1)} - \frac{c^2}{2m} \varepsilon^2 t_{(0)}
  \right)}
  \nonumber\\
  &+
F_-\left( x_{(0)} +\frac{c}{m}\sqrt{2m\varepsilon}\ t_{(0)}   \right)
  e^{-i\left( \varepsilon t_{(1)} + \frac{\sqrt{2m\varepsilon}}{c} x_{(1)} - \frac{c^2}{2m}\varepsilon^2  t_{(0)}
  \right)},
  \label{solM1}
\end{align}
where we have written a $\varepsilon$ sub-index to indicate the dependence on the parameter $\varepsilon$. Alternatively, we can identify the solution using ${p}$ to refer to the two eigenvalues ${\pm p}$ of the momentum operator $-i\partial_{x_{(1)}}$ for the two components, with $\varepsilon=c^2/(2m){p}^2$, and write
\begin{align}
	\Psi_{p}(t_{(0)},t_{(1)},x_{(0)},x_{(1)}) &=
	 F_+\left( x_{(0)} - \frac{c^2}{m}{p} t_{(0)}   \right)
	e^{-i\left( \frac{c^2}{2m}{p}^2 t_{(1)} - {p} x_{(1)} 
		- \frac{c^6}{8m^3} {p}^4 t_{(0)}
		\right)}
	\nonumber\\
	&+
	 F_-\left( x_{(0)} + \frac{c^2}{m}{p} t_{(0)}   \right)
	e^{-i\left( \frac{c^2}{2m}{p}^2 t_{(1)} + {p} x_{(1)} 
		- \frac{c^6}{8m^3} {p}^4 t_{(0)}
		\right)},
	\label{solM1b}
\end{align}

This method can be repeated to solve the system of Schr\"odinger equations for any $M>1$. The constraint for $k=M$ yields a   equation for the dependence on $t_{(M)}, x_{(M)}$ which is second order in $x_{(M)}$ and which can be solved by separation of variables, yielding left and right travelling waves in $t_{(M)}, x_{(M)}$ . Each of the two solutions can then be substituted in the equation for $k=M-1$, and one gets a first order PDE in $t_{(M-1)}, x_{(M-1)}$, which can be solved by the method of characteristics. The obtained solutions can be substituted in the equation for $k=M-2$ which is again of first order in $t_{(M-2)}, x_{(M-2)}$, and the procedure can be iterated until we reach the equation for $k=0$.

\subsection*{Projective phase}

We will discuss here the transformation properties of the above  wave functions and
 Schr\"odinger equations and the associated projective phases
 under the post-Galilean transformations.
 For simplicity we will work again in $d=1$. We will discuss the case of expansions in detail and summarize the results for the other transformations. Since we have $d=1$ we do not consider rotations and generalized rotations. Also, it should be taken into account that the finite transformations presented below are those of the individual generators. 

\begin{enumerate}
\item\textbf{Expansions $C$.}
The expansions for $M=1$ define a vector field in the cotangent manifold,
 with coordinates $(t_{(0)},x_{(0)}, t_{(1)},x_{(1)},E^{(0)},E^{(1)},p^{(0)},p^{(1)})$,
given by
\begin{align}
X^C &= t_{(0)}x_{(0)} \frac{\partial}{\partial x_{(1)}} + t_{(0)}^2 \frac{\partial}{\partial t_{(1)}} +
\left(-t_{(0)} p^{(1)} + \frac{m}{c^2} x_{(0)} \right) \frac{\partial}{\partial p^{(0)}}
\nonumber\\
&+\left(
-2E^{(1)} t_{(0)} + p^{(1)} x_{(0)}
\right)\frac{\partial}{\partial E^{(0)}},
\label{XC}
\end{align}	
which, after a trivial integration, yields the finite expansions
\begin{align}
	\hat{x}_{(0)} = x_{(0)}, &\quad  \hat{p}^{(0)} =p^{(0)} + \mu \left(-t_{(0)} p^{(1)} + \frac{m}{c^2} x_{(0)} \right),\\
	\hat{x}_{(1)} = x_{(1)} + \mu t_{(0)}x_{(0)} , &\quad \hat{p}^{(1)} = p^{(1)},\\
	\hat{t}_{(0)} = t_{(0)}, &\quad  \hat{E}^{(0)} = E^{(0)} + \mu \left(
	-2E^{(1)} t_{(0)} + p^{(1)} x_{(0)}
	\right),\\
	\hat{t}_{(1)} = t_{(1)} + \mu t_{(0)}^2, &\quad \hat{E}^{(1)} = E^{(1)}.
\end{align}
Notice that $t_{(0)}$ and $x_{(0)}$ do not transform under $C$, while for the standard
Schr\"odinger expansions ($M=0$) one has
\begin{equation}
\hat{t}_{(0)} = \frac{t_{(0)}}{1-\mu t_{(0)}},\quad 
\hat{x}_{(0)} = \frac{x_{(0)}}{1-\mu t_{(0)}}.
\end{equation}

Let $\Psi$ be a solution to (\ref{Sch1}) and (\ref{Sch2}), and let $\hat\Psi$ be the transformed solution under a  finite $C$ transformation.
Assume that $\hat\Psi$ and $\Psi$ are related by
\begin{equation}
\hat\Psi(\hat{t}_{(0)},\hat{t}_{(1)},\hat{x}_{(0)},\hat{x}_{(1)}) =
e^{i\varphi(t_{(0)},t_{(1)},x_{(0)},x_{(1)})} \Psi(t_{(0)},t_{(1)},x_{(0)},x_{(1)}).
\end{equation}

If we demand
 $\hat\Psi$ to satisfy (\ref{Sch1}) and (\ref{Sch2}) in the transformed coordinates one gets,
 after some algebra, the following set of PDE for $\varphi$
\begin{equation}
\frac{\partial\varphi}{\partial t_{(1)}} =\frac{\partial\varphi}{\partial x_{(1)}}  
=\frac{\partial\varphi}{\partial t_{(0)}} = 0,\quad 
\frac{\partial\varphi}{\partial x_{(0)}} - \mu\frac{m}{c^2} x_{(0)}=0,  
\end{equation}
whose solution is
\begin{equation}
\varphi(t_{(0)},t_{(1)},x_{(0)},x_{(1)}) = \mu \frac{m}{2c^2} x_{(0)}^2 + \text{constant}.
\end{equation}
Since the Jacobian 
\begin{equation}
\frac{\partial(\hat{x}_{(1)},\hat{x}_{(0)})}{\partial({x}_{(1)},{x}_{(0)})}=
\left|
\begin{array}{cc}
 1 & \mu t_{(0)} \\ 0 & 1
\end{array}
\right|=1
\end{equation}
is trivial, an imaginary part for the constant is not needed to compensate for a change in the measure in $x_{(0)},x_{(1)}$ space, and we can take it equal to zero.
Again, this is different from what happens in the $M=0$ case, for which the projective phase acquires an imaginary part,
\begin{equation}
\varphi(t_{(0)},x_{(0)}) = \frac{1}{2}\frac{m\mu}{1-\mu t_{(0)}} x_{(0)}^2 
-\frac{i}{2} \log|1-\mu t_{(0)}|.
\end{equation}
One has then
\begin{equation}
|\hat\Psi|^2 = |1-\mu t_{(0)}||\Psi|^2,
\end{equation}
so that $|\hat\Psi|^2 \dif\hat{x}_{(0)} = |\Psi|^2 \dif x_{(0)}$, as desired.

Returning to the $M=1$ expansions, we have
\begin{equation}
\hat\Psi(\hat{t}_{(0)},\hat{t}_{(1)},\hat{x}_{(0)},\hat{x}_{(1)}) =
e^{i \mu \frac{m}{2c^2} x_{(0)}^2} \Psi(t_{(0)},t_{(1)},x_{(0)},x_{(1)}).
\label{PFCM1}
\end{equation}

Notice that  the projective phase can also be obtained by iteration of the infinitesimal $\delta F$ corresponding to expansions,
\begin{equation}
\varphi = \sum_{n=1}^\infty \frac{1}{n!}\delta^nF,
\label{phiF}
\end{equation}
with $\delta^n F = \delta (\delta^{n-1}F)$. Since  $\delta F=m/(2c^2)\mu x_{(0)}^2$ and  $\delta x_{(0)}=0$ for expansions in the $M=1$ case, the contributions are null after the lineal one.  For a general discussion about projective phase, central extensions and invariance up to a total derivative of the Lagrangian see \cite{levy1969group,marmo1988quasi,Silagadze:2011yj}.

As a check, let us assume that $\Psi$  is a solution of (\ref{Sch1}), (\ref{Sch2}) given by  (\ref{solM1}), and let us prove that then $\hat\Psi$ has the same form. Using (\ref{solM1}), the right-hand side of (\ref{PFCM1}) is
\begin{align}
 & 
F_+\left( x_{(0)} - \frac{c}{m}\sqrt{2m\varepsilon}\ t_{(0)}   \right)
e^{-i\left( \varepsilon t_{(1)} - \frac{\sqrt{2m\varepsilon}}{c} x_{(1)} - \frac{c^2}{2m} \varepsilon^2 t_{(0)}
	\right)}e^{i \mu \frac{m}{2c^2} x_{(0)}^2}
\nonumber\\
&+
F_-\left( x_{(0)} +\frac{c}{m}\sqrt{2m\varepsilon}\ t_{(0)}   \right)
e^{-i\left( \varepsilon t_{(1)} + \frac{\sqrt{2m\varepsilon}}{c} x_{(1)} - \frac{c^2}{2m}\varepsilon^2 {t}_{(0)}
	\right)}e^{i \mu \frac{m}{2c^2} x_{(0)}^2},\nonumber\\
&=
F_+\left( \hat{x}_{(0)} - \frac{c}{m}\sqrt{2m\varepsilon}\ \hat{t}_{(0)}   \right)
e^{-i\left( \varepsilon t_{(1)} - \frac{\sqrt{2m\varepsilon}}{c} x_{(1)} - \frac{c^2}{2m}\varepsilon^2 \hat{t}_{(0)}
	\right)}e^{i \mu \frac{m}{2c^2} \hat{x}_{(0)}^2}
\nonumber\\
&+
F_-\left( \hat{x}_{(0)} +\frac{c}{m}\sqrt{2m\varepsilon}\ \hat{t}_{(0)}   \right)
e^{-i\left( \varepsilon t_{(1)} + \frac{\sqrt{2m\varepsilon}}{c} x_{(1)} - \frac{c^2}{2m} \varepsilon^2 \hat{t}_{(0)}
	\right)}e^{i \mu \frac{m}{2c^2} \hat{x}_{(0)}^2},
\end{align}
where we have used that $\hat{t}_{(0)}=t_{(0)}$, $\hat{x}_{(0)} = x_{(0)}$. Expressing now $t_{(1)}$ and $x_{(1)}$ in terms of the transformed variables one gets
\begin{align}
	\varepsilon t_{(1)} \mp \frac{\sqrt{2m\varepsilon}}{c} x_{(1)} &=
	 \varepsilon \hat{t}_{(1)} \mp \frac{\sqrt{2m\varepsilon}}{c} \hat{x}_{(1)} 
	 -\mu \left(
	 \varepsilon \hat{t}_{(0)}^2 \mp \frac{\sqrt{2m\varepsilon}}{c} \hat{t}_{(0)}
	 \hat{x}_{(0)}
	 \right).   
\end{align}
The extra terms can be combined with  the one with $\hat{x}_{(0)}^2$ and one gets the perfect square
\begin{equation}
-i\mu \frac{m}{2c^2} \left(
\hat{x}_{(0)} \mp \frac{c}{m}\sqrt{2m\varepsilon} \hat{t}_{(0)}
\right)^2,
\end{equation}
whose exponential can then be absorbed into $F_{\pm}$ to yield the corresponding $\hat{F}_\pm$ for $\hat\Psi$, giving it the same form as in (\ref{solM1}) but in transformed variables. Notice that the parameter $\varepsilon$ does not change when going to the transformed coordinates, since for a given solution it corresponds to the value of $E^{(1)}$, and under the expansions one has $\hat{E}^{(1)}=E^{(1)}$. This is not, however, true of some of the other transformations and the change of $E^{(1)}$, or of $\frac{c^2}{2m}p^{(1)2}$, must be taken into account.

\item\textbf{Boosts $B^{(0)}$.}
The finite transformations for parameter $v_{(0)}=v$ are in this case
\begin{align}
	\hat{x}_{(0)} &= x_{(0)}+ v t_{(0)}, \\
	 \hat{p}^{(0)} &=p^{(0)} + v E^{(1)} +\frac{1}{2}v^2 p^{(1)} + \frac{1}{6}\frac{m}{c^2}v^3,\\
	\hat{x}_{(1)} &= x_{(1)} + v t_{(1)}+\frac{1}{2}v^2 x_{(0)}+\frac{1}{6}v^3 t_{(0)}, 
	\\
	\hat{p}^{(1)} &= p^{(1)} + \frac{m}{c^2}v,\\
	\hat{t}_{(0)} &= t_{(0)},\\ 
	 \hat{E}^{(0)} &= E^{(0)} + v p^{(0)} + \frac{1}{2}v^2 E^{(1)} + \frac{1}{6} v^3 p^{(1)} + \frac{1}{24}\frac{m}{c^2}v^4,\\
	\hat{t}_{(1)} &= t_{(1)}+v x_{(0)} + \frac{1}{2}v^2 t_{(0)}, \\
	\hat{E}^{(1)} &= E^{(1)}+ v p^{(1)} + \frac{1}{2}\frac{m}{c^2}v^2,
\end{align}
and the projective phase is 
\begin{equation}
\varphi=\frac{m}{c^2}\left(
v x_{(1)} + \frac{1}{2} v^2 t_{(1)} + \frac{1}{6} v^3 x_{(0)} + \frac{1}{24}v^4 t_{(0)}
\right).
\end{equation}

\item\textbf{Generalized boosts $B^{(1)}$.}
The finite transformations corresponding to parameter $v_{(1)}=v$ are
\begin{align}
	\hat{x}_{(0)} = x_{(0)}, &\quad  \hat{p}^{(0)} =p^{(0)} + \frac{m}{c^2}v,\\
	\hat{x}_{(1)} = x_{(1)} + v t_{(0)}, &\quad \hat{p}^{(1)} = p^{(1)},\\
	\hat{t}_{(0)} = t_{(0)}, &\quad  \hat{E}^{(0)} = E^{(0)} + v p^{(1)},\\
	\hat{t}_{(1)} = t_{(1)}, &\quad \hat{E}^{(1)} = E^{(1)}.
\end{align}
The projective phase is in this case
\begin{equation}
\varphi=\frac{m}{c^2}v x_{(0)}.
\end{equation}
\item\textbf{Dilatations $D$.}
The finite transformations for parameter $\lambda$ are
\begin{align}
	\hat{x}_{(0)} = e^{\lambda} x_{(0)}, &\quad  \hat{p}^{(0)} =e^{-\lambda}p^{(0)} ,\\
	\hat{x}_{(1)} = e^{\lambda/3}x_{(1)}, &\quad  \hat{p}^{(1)} = e^{-\lambda/3} p^{(1)},\\
	\hat{t}_{(0)} = e^{4\lambda/3}t_{(0)}, &\quad  \hat{E}^{(0)} = e^{-4\lambda/3}E^{(0)} ,\\
	\hat{t}_{(1)} = e^{2\lambda/3}t_{(1)}, &\quad  \hat{E}^{(1)} =e^{-2\lambda/3}t_{(1)} E^{(1)}.
\end{align}
The projective phase is a constant which, however, cannot be taken as zero and in fact must be imaginary
$\varphi=i\frac{2}{3}\lambda$,
to compensate for the change in the measure in $x_{(0)}, x_{(1)}$ space,
$\dif\hat{x}_{(0)}\dif\hat{x}_{(1)} = e^{4\lambda/3}\dif x_{(0)}\dif x_{(1)}$.

\item\textbf{Generalized dilatations $D^{(1)}$.}
The finite transformations corresponding to parameter $\lambda_{(1)}=\lambda$ are
\begin{align}
	\hat{x}_{(0)} = x_{(0)}, &\quad  \hat{p}^{(0)} =p^{(0)} -\frac{1}{2}\lambda p^{(1)},\\
	\hat{x}_{(1)} = x_{(1)} + \frac{1}{2}\lambda x_{(0)}, &\quad \hat{p}^{(1)} = p^{(1)},\\
	\hat{t}_{(0)} = t_{(0)}, &\quad  \hat{E}^{(0)} = E^{(0)} - \lambda E^{(1)},\\
	\hat{t}_{(1)} = t_{(1)}+ \lambda t_{(0)}, &\quad \hat{E}^{(1)} = E^{(1)}.
\end{align}
and the projective phase is trivial. No constant imaginary part for $\varphi$ is needed, since
$\dif\hat{x}_{(0)}\dif\hat{x}_{(1)} = \dif x_{(0)}\dif x_{(1)}$ in this case.

\item\textbf{Time shifts and space translations $H^{(0)}$, $H^{(1)}$, $P^{(0)}$, $P^{(1)}$.}
The finite transformations are just shifts in the corresponding variables, the momenta do not transform and the projective phase can be chosen as zero in all the cases.
\end{enumerate}

\subsection*{Schr\"odinger equation with higher derivatives}

The way that we have constructed the Schr\"odinger equation of our system corresponds to what is known as weak quantization, where  the constraints are imposed as operators on the states of the system. One can also consider a reduced space quantization, in which the gauge invariance is broken and a Hamiltonian in then computed. We can do this for our action (\ref{S2}), disregarding the total derivative, by imposing the two gauge conditions \cite{Gomis:2019sqv}
\begin{equation}
t_{(0)} = c^{-2}t_{(1)}=\tau,
\label{GF}
\end{equation}
whereby one obtains the gauge fixed action
\begin{equation}
S^*_{(2)} = \frac{m}{c^2}\int\dif t \left(
\dot{x}_{(0)}^a \dot{x}_{(1)a} - \frac{c^2}{2} \dot{x}_{(0)}^a \dot{x}_{(0)a} + \frac{1}{8} (\dot{x}_{(0)}^a \dot{x}_{(0)a})^2
\right) .
\end{equation}
From this one can  compute the Hamiltonian
\begin{equation}
H_{(2)}^* = \frac{c^2}{m} p_{(0)a} p_{(1)}^a  + \frac{1}{2} \frac{c^4}{m} p_{(1)a}p_{(1)}^a - \frac{1}{8} \frac{c^6}{m^3} ( p_{(1)a}p_{(1)}^a)^2,
\end{equation}
and write down the corresponding time-dependent Schr\"odinger equation for $\Psi(t,x_{(0)},x_{(1)})$
\begin{equation}
i \frac{\partial\Psi}{\partial t} =  - \frac{c^2}{m} \frac{\partial^2\Psi}{\partial{x_{(0)a}} \partial{x_{(1)}^a}}
- \frac{1}{2}\frac{c^4}{m} \frac{\partial^2\Psi}{\partial{x_{(1)a}} \partial{x_{(1)}^a}} 
- \frac{1}{8}\frac{c^6}{m^3} \frac{\partial^4\Psi}{\partial{x_{(1)a}}^2 \partial{x_{(1)}^a}^2},
\label{Sch4}
\end{equation}
 which is of fourth order. We can also obtain an equation of this order by plugging (\ref{Sch2}) into (\ref{Sch1}),
 \begin{equation}
 i \frac{\partial\Psi}{\partial t_{(0)}} =  - \frac{c^2}{m} \frac{\partial^2\Psi}{\partial{x_{(0)a}} \partial{x_{(1)}^a}}
 - \frac{1}{8}\frac{c^6}{m^3} \frac{\partial^4\Psi}{\partial{x_{(1)a}}^2 \partial{x_{(1)}^a}^2},
 \label{Sch5}
 \end{equation}
which is also of fourth order and with the same coefficients, but lacks the term with second derivatives in $x_{(1)}$. One can conclude then that the system described by action (\ref{S2}) is one for which the two quantization procedures yield different results. Nevertheless, one should notice that the wave function obtained by using the gauge fixing (\ref{GF}) in (\ref{solM1b}), which by construction is a solution of (\ref{Sch5}) with $t_{(0)}=t$, is also a solution of (\ref{Sch4}).

 \section{Conclusions and outlook}
 \label{conclusions}

 In this paper we have constructed the most general point symmetry transformations of 
 the post Galilean actions  \cite{Gomis:2019sqv}  that can be obtained from the Minkowskian action for a massive particle. The  algebras  obtained are 
 generalizations of the ordinary Schr\"odinger algebra~\cite{Niederer:1972zz,Hagen:1972pd}. Besides the generalized Galilean transformations, they contain dilatations, $D$, generalized dilatations $D^{(k)}$ and expansions $C$.
 The algebras are different from extensions of Galilean conformal algebras with 
 dynamical exponent $z=2/N$, with $N$ positive integer, since these contain an $sl(2,\R)$
 subalgebra \cite{henkel1997local,negro1997nonrelativistic}.

 Using a weak quantization procedure,
 we have  introduced an Schr\"odinger equation for the post-Galilean particle that consists of $M+1$ partial differential equations, up to second order in derivatives, for a wave function living in a generalized space. Like the case of ordinary Schr\"odinger equation, the wave function 
 supports a ray representation of the symmetry group, and we have calculated the projective 
 phase for each transformation. 
 The symmetries of generalized Schr\"odinger  equations in this paper are different from the symmetries of the higher order Schr\"odinger  equations \cite{Gomis:2011dw}.
 
 If we consider the reduced  space quantization the 
corresponding Schr\"odinger equation is a single differential equation of fourth order.
The two procedures of quantization do not coincide in general.
Further investigation of the difference between the Schr\"odinger equations obtained from the weak and reduced space quantizations,  and the generalization of this fact for the actions $S_{(M+1)}$,  will also be the subject of future work.

It will be interesting to study the relation of the higher order Schr\"odinger
equation with the expansion up to order  $v^2/c^2$ of the square root
of the Klein-Gordon equation, see for example \cite{sucher1963relativistic,fiziev1985relativistic},
\begin{equation}
 i \frac{\partial\Psi}{\partial t} =  \sqrt{- c^2 \frac{\partial^2}{\partial{x^{a}} \partial{x_a}}+m^2 c^4}\,\Psi,
 \label{Squareroot}
 \end{equation}
and if it is possible to introduce interaction terms in the new Schr\"odinger equation that we have found.

{{}
Finally, another topic of interest for future research is the computation of quadratic invariants under stability subgroups of the generalized Schr\"odinger algebras and their use  to construct associated space-times, as done for instance in \cite{Brugues:2006yd} to obtain  pp-wave metrics from Newton-Hooke algebras.
}

 \section*{Acknowledgements}
 We would like to thank Eric Bergshoeff, Gary Gibbons, Axel Kleinschmidt, Patricio Salgado-Rebolledo and Paul Townsend for their comments on the paper. 
  {{}
 We would also like to thank the anonymous reviewer of the first version of this paper for his/her insightful observations.
}
 
 The work of CB has been partially supported by Generalitat de Catalunya through project 2017 SGR 872 and by
 Spanish national project  DOVELAR-IRI, RTI2018-096001-B-C32. JG has been supported in part by MINECO FPA2016-76005-C2-1-P and Consolider CPAN, and by the Spanish government (MINECO/FEDER) under project MDM-2014-0369 of ICCUB (Unidad de Excelencia Mar\'{i}a de Maeztu).

 \bibliographystyle{JHEP}
 \bibliography{postnewtonian}

\vfill 
\appendix
 
\section{Invariance of the contraints under  generalized boosts and rotations}
\label{inv_constraints}
 Under the boosts in (\ref{Ep_trans})  the variation of the constraints (\ref{constraintsM})  is 
\begin{align}
\delta \phi^{(k)} &=  - \frac{m}{c^{2M}} \sum_{j=0}^{M-k} v_{(j)}^a p_a^{(k+j)}\label{var_I}\\
& + \frac{1}{2} \sum_{l=k}^{M-1} \sum_{j=0}^{M-l-1} v_{(j)}^a E^{(l+j+1)} p_a^{(M+k-l)}\label{var_IV}\\
& + \frac{1}{2}\frac{m}{c^{2M}} \sum_{l=k}^M v_{(M-l)}^a p_a^{(M+k-l)}\label{var_II}\\
& + \frac{1}{2} \sum_{l=k+1}^M \sum_{j=0}^{l-k-1} v_{(j)}^a  p_a^{(l)}  E^{(M+k-l+j+1)}\label{var_V}\\
& + \frac{1}{2} \frac{m}{c^{2M}} \sum_{l=k}^M v_{(l-k)}^a p_a^{(l)}\label{var_III}\\
& - \frac{1}{2} \sum_{l=k}^{M-1} \sum_{j=0}^{M-l-1} v_{(j)}^a p_a^{(l+j+1)} E^{(M+k-l)}\label{var_VI}\\
& - \frac{1}{2} \sum_{l=k}^{M-1} \sum_{j=0}^{l-k} v_{(j)}^a E^{(l+1)} p_a^{(M+k-l+j)}\label{var_VII}.
\end{align}
Sums in (\ref{var_II}) and (\ref{var_III}) can be brought to the same form as in (\ref{var_I}) under the changes of indexes $l\to  j = M-l$ and $l \to j =l-k$, respectively, and hence cancel overall. Performing the double change of indexes $(l,j)\to (\hat l,\hat j) =(l-j,j)$ in (\ref{var_VII}) brings it to the same form as in (\ref{var_IV}), and hence both terms cancel each other. Finally, after  performing the change
$(l,j)\to (\hat l,\hat j) =(l-j-1,j)$ in (\ref{var_V}) one sees that it exactly cancels (\ref{var_VI}), and hence the constraint is invariant under boosts.

For the rotations, for which $\delta E^{(k)}=0$, one gets
\begin{align}
\delta \phi^{(k)} &=\frac{1}{2} \sum_{l=k}^M \sum_{j=0}^{M-l}  \omega_{(j)ab}\ p^{(l+j)b} p^{(M+k-l)a}  
 + \frac{1}{2}\sum_{l=k}^M \sum_{j=0}^{l-k} \omega_{(j)ab}\ p^{(l)a} p^{(M+k-l+j)b}.
 \end{align} 
 Performing the  change of indexes $(l,j)\to (\hat l,\hat j) =(l-j,j)$ in the second double sum and using the skew-symmetry of the $\omega^{(k)}$ one sees that this cancels the first double sum, and hence $\delta \phi^{(k)}=0$ under rotations as well.
 
 \section{Transformation of the canonical variables under the $M$th level symmetry generators}
 \label{TVC}
 Given $M\geq 0$ we compute the transformation of the canonical variables $x_{(j)}$, $p^{(j)}$, $t_{(j)}$ and $E^{(j)}$ for $j=0,1,\ldots,M$. In all the following results it should be understood that if a canonical variable appears with an index outside of the range $[0,M]$ the corresponding result should be treated as zero.
\subsection*{Dilatation $D$.}
\begin{align}
\delta^D x_{(j)}^a &= \left\{ x_{(j)}^a,D  \right\} = \frac{2M+1-2j}{2M+1} x_{(j)}^a,\\
\delta^D p^{(j)}_a &= \left\{ p^{(j)}_a,D  \right\} = - \frac{2M+1-2j}{2M+1} p_a^{(j)},\\
\delta^D t_{(j)} &= \left\{ t_{(j)},D  \right\} = \frac{2M+2-2j}{2M+1} t_{(j)},\\
\delta^D E^{(j)} &= \left\{ E^{(j)},D  \right\} = - \frac{2M+2-2j}{2M+1} E^{(j)}.
\end{align}
\subsection*{Generalized dilatations $D^{(k)}$, $k=1,\ldots,M$.}
\begin{align}
\delta^{D,k} x_{(j)}^a &= \left\{ x_{(j)}^a,D^{(k)}  \right\} = \frac{2(M-j)+1}{2(M+1-k)} x_{(j-k)}^a,\\
\delta^{D,k} p^{(j)}_a &= \left\{ p^{(j)}_a,D^{(k)}  \right\} = - \frac{2(M-k-j)+1}{2(M+1-k)} p_a^{(j+k)},\\
\delta^{D,k} t_{(j)} &= \left\{ t_{(j)},D^{(k)}  \right\} = \frac{M+1-j}{M+1-k} t_{(j-k)},\\
\delta^{D,k} E^{(j)} &= \left\{ E^{(j)},D^{(k)}  \right\} = - \frac{M+1-k-j}{M+1-k} E^{(j+k)}.
\end{align}

\subsection*{Expansions $C$.}
\begin{align}
\delta^C x_{(j)}^a &= \left\{ x_{(j)}^a,C  \right\} = \delta^M_j t_{(0)} x_{(0)}^a,\\
\delta^C p^{(j)}_a &= \left\{ p^{(j)}_a,C  \right\} =\delta^j_0 \left(  -t_{(0)}p^{(M)}_a + \frac{m}{c^{2M}} x_{(0)a}      \right),\\
\delta^C t_{(j)} &= \left\{ t_{(j)},C \right\} = \delta^M_j t_{(0)}^2,\\
\delta^C E^{(j)} &= \left\{ E^{(j)},C  \right\} = \delta^j_0 \left( -2E^{(M)} t_{(0)} + p^{(M)}_b x_{(0)}^b \right).
\end{align}

\subsection*{Boosts and generalized boosts $B^{(k)}$, $k=0,\ldots,M$.}
\begin{align}
\delta^{B,k}_a x_{(j)}^b &= \left\{ x_{(j)}^b,B^{(k)}_a  \right\} = \delta_a^b t_{(j-k)},\\
\delta^{B,k}_a p^{(j)}_b &= \left\{ p^{(j)}_b,B^{(k)}_a  \right\} = \delta_{ab} E^{(j+k+1)}+ \frac{m}{c^{2M}} \delta_{ab} \delta^j_{M-k},\\
\delta^{B,k}_a t_{(j)} &= \left\{ t_{(j)},B^{(k)}_a  \right\} = x_{(j-k-1)a},\\
\delta^{B,k}_a E^{(j)} &= \left\{ E^{(j)},B^{(k)}_a  \right\} = p^{(j+k)}_a.
\end{align}

\subsection*{Time shifts $H^{(k)}$, $k=0,\ldots,M$.}
\begin{align}
\delta^{H,k} x_{(j)}^a &= \left\{ x_{(j)}^a,H^{(k)}  \right\} = 0,\\
\delta^{H,k} p^{(j)}_a &= \left\{ p^{(j)}_a,H^{(k)}  \right\} =0,\\
\delta^{H,k} t_{(j)} &= \left\{ t_{(j)},H^{(k)}  \right\} = \delta_j^k,\\
\delta^{H,k} E^{(j)} &= \left\{ E^{(j)},H^{(k)}  \right\} =0.
\end{align}

\subsection*{Space translations $P^{(k)}$, $k=0,\ldots,M$.}
\begin{align}
\delta^{P,k}_a x_{(j)}^b &= \left\{ x_{(j)}^b,P^{(k)}_a  \right\} = \delta_j^k \delta_a^b,\\
\delta^{P,k}_a p^{(j)}_b &= \left\{ p^{(j)}_b,P^{(k)}_a  \right\} =0,\\
\delta^{P,k}_a t_{(j)} &= \left\{ t_{(j)},P^{(k)}_a  \right\} = 0,\\
\delta^{P,k}_a E^{(j)} &= \left\{ E^{(j)},P^{(k)}_a  \right\} =0.
\end{align}

\subsection*{Rotations and generalized rotations $J^{(k)}$, $k=0,\ldots,M$.}
\begin{align}
\delta^{J,k}_{ab} x_{(j)}^c &= \left\{ x_{(j)}^c,J^{(k)}_{ab}  \right\} = \delta_a^c x_{(j-k)b}- \delta_b^c x_{(j-k)a},\\
\delta^{J,k}_{ab} p^{(j)}_c &= \left\{ p^{(j)}_c,J^{(k)}_{ab}  \right\} = -\delta_{bc} p^{(j+k)}_a+  \delta_{ac} p^{(j+k)}_b,\\
\delta^{J,k}_{ab} t_{(j)} &= \left\{ t_{(j)},J^{(k)}_{ab}  \right\} = 0,\\
\delta^{J,k}_{ab} E^{(j)} &= \left\{ E^{(j)},J^{(k)}_{ab}  \right\} = 0.
\end{align}

The transformations of the $p^{(j)}$ and $E^{(j)}$ under boosts and rotations agree with those proposed in Section \ref{can_act}.

\section{Invariance of the constraints under the full set of transformations}
\label{IVC}
Here we show that the proposed constraints  (\ref{constraintsM}) are invariant (or weakly invariant in some cases) under the full set of transformations given by (\ref{fullgen}). Since for boosts and rotations the transformations given by the  generators correspond to those 
of (\ref{Ep_trans}), for which the invariance of the constraints is proved in Appendix \ref{inv_constraints}, we will deal here only with the rest of generators, that is $D$, $C$, $D^{(k)}$, $H^{(k)}$ and $P^{(k)}$. We will consider the transformations under the individual generators without multiplying by the corresponding parameters, using the results in Appendix (\ref{TVC}). 

\subsection*{Dilatations $D$.}
The terms which add up to form $\phi^{(k)}$ transform homogeneously under $D$ and one has
\begin{equation}
\delta^D\phi^{(k)} = - \frac{2M+2-2k}{2M+1} \phi_{(k)} \approx 0,\quad  k=0,1,\ldots,M,
\end{equation}
where $\approx$ means equality on the constraints manifold, so that the constraints are weakly invariant under dilatations.

\subsection*{Expansions $C$.}
For expansions one has
\begin{equation}
\delta^C\phi^{(k)} = \begin{cases}
-2 t_{(0)} \phi^{(M)} & \text{for $k=0$},\\
0 & \text{for $k=1,\ldots,M$},
\end{cases}
\end{equation}
and the constraints are also weakly invariant.

\subsection*{Generalized dilatations $D^{(j)}$, $j=1,\ldots,M$.}
Under generalized dilatations one has
\begin{equation}
\delta^{D,j}\phi^{(k)} = 
\begin{cases}
-\frac{M+1-j-k}{M+1-j} \phi^{(k+j)} & \text{if $k+j\leq M$},\\
0 & \text{otherwise},
\end{cases}
\end{equation}
and we have weakly invariance.

\subsection*{Time shifts and space translations, $H^{(j)}$, $P^{(j)}$, $j=0,\ldots,M$.}
The $P^{(l)}$ and $E^{(l)}$ are invariant under all these transformations, and hence so are the $\phi^{(k)}$,
\begin{equation}
\delta^{H,j}\phi^{(k)} = \delta^{P,j}_a\phi^{(k)}=0, \quad j,k=0,1,\ldots,M.
\end{equation}

\end{document}